# t factor: A metric for measuring impact on Twitter

**Lutz Bornmann[1*] & Robin Haunschild[2]**
[1]Administrative Headquarters of the Max Planck Society
Division for Science and Innovation Studies
Munich, GERMANY
[2]Max Planck Institute for Solid State Research
Stuttgart, GERMANY.
e-mail: *bornmann@gv.mpg.de (corresponding author); R.Haunschild@fkf.mpg.de

## ABSTRACT

*Based on the definition of the well-known h index we propose a t factor for measuring the impact of publications (and other entities) on Twitter. The new index combines tweet and retweet data in a balanced way, whereby retweets are seen as data reflecting the impact of initial tweets. The t factor is defined as follows: a unit (single publication, journal, researcher, research group etc.) has factor t if t of its $N_t$ tweets have at least t retweets each and the other ($N_t$-t) tweets have ≤t retweets each.*

## INTRODUCTION

Alternative metrics (altmetrics) has been proposed to measure the broad impact of science (Bornmann 2014). Here, saves, tweets, shares, likes, and recommends, tags, posts and other activities on social media are counted to measure impact. Twitter is one of the most important sources of altmetric data. As a rule, success on Twitter is measured by two options: (i) the number of tweets, or (ii) the number of followers: For example, Haustein et al. (2014) provide evidence about how often Twitter is used to disseminate information about papers in biomedicine by using their number of mentions in tweets. You (2014) presents the top 20 of the 50 most followed scientists on Twitter (see also Darling et al. 2013).

We would like to propose a third option to measure Twitter impact which considers re-tweets. Since re-tweets are simple copies of original tweets, they reflect the interest of a follower in a tweet. The more often a tweet is re-tweeted the more the content of the initial tweets is of general interest. Thus, we propose to consider re-tweets as impact data of initial tweets whereby re-tweet data of single tweets measure their success: Our approach combines the number of tweets (for a publication, researcher etc.) and the number of re-tweets for the single tweets using the formula of the h index (Hirsch 2005).

## LITERATURE OVERVIEW

Twitter is "the best known micro blogging application, which has achieved rapid and well-publicized growth. Though it began as a way for users to answer the question, 'What are you doing?', its ease of use, enforced brevity, and wide reach has encouraged users to employ Twitter for more serious tasks, as well" (Priem & Hemminger 2010). In the Snowball Metrics Recipe Book – an initiative owned by research-intensive universities to agree on methodologies for institutional benchmarking – Colledge (2014) mention Twitter counts as one of the metrics which can be used to measure social activities. Elsevier is also involved in the Snowball Metrics initiative. Scopus (Elsevier) does not only show Twitter counts for single publications, but also normalized Twitter counts. It presents in which percentile rank a publication belongs compared to similar publications. However, previous research could not reveal what Twitter counts really measure; the correlation with traditional citation counts is negligible (Bornmann 2015).

"Quite frequently, Twitter users directly quote other peoples' tweets. Tweets are either copied completely, or users copy parts of an existing tweet and add their own comment. In many cases, the users also mention the original author – this clearly resembles citation practices in scientific communication. Because these copied tweets have often been labeled as 'Re-tweets' or 'RTs' by Twitter users, Twitter has established re-tweeting as a genuine Twitter functionality" (Weller und Peters 2012, 212; see also Zubiaga et al. 2014). According to the results of Holmberg und Thelwall (2014) up to a third of tweets may be re-tweets. Although Weller and Peters (2012) suggested that re-tweets should be named as internal citations (which should be differentiated from other tweets), most previous investigations of Twitter counts have counted tweets without differentiating between initial tweets and re-tweets.

## METHODS
### Dataset used
We would like to demonstrate the calculation of the t factor on the level of a single publication. In Scopus, we searched for all tweets and re-tweets of the paper "An index to quantify an individual's scientific research output" published by Hirsch (2005). The total of 69 tweets published between June 17, 2011 and June 16, 2015 were downloaded. Since Scopus does not differentiate between tweets and re-tweets, we categorized those tweets as re-tweets which contain the same content as a tweet published subsequently.

### Definition of the t factor
We propose to combine the number of tweets (for a publication, researcher etc.) and the number of re-tweets for the single tweets using the formula of the h index (Hirsch 2005). The formula is based on two types of information: (i) the number of papers ($N_p$) published over *n* years and (ii) the number of citations for each paper. "A scientist has index *h* if *h* of his or her $N_p$ papers have at least *h* citations each and the other ($N_p$-*h*) papers have ≤*h* citations each" (p. 16569).[1] The h index is a very popular metric for measuring research performance, which has been included in databases like Web of Science (WoS, Thomson Reuters) and Scopus. However, it has been criticized that the index can only be used if the scientists (which are compared) have published in the same field, time period and are in a similar age.

---

[1] The idea of combining tweets and retweets in this specific way was introduced at http://limitednews.com.au/2012/12/the-zap-index-are-you-better-or-worser-at-twitter/

The corresponding Twitter (t) factor combines the following information: (i) the number of tweets ($N_t$) published over a certain time period and (ii) the number of retweets for each tweet. The definition of the Twitter factor is: A unit (single publication, journal, researcher, research group etc.) has Twitter factor *t* if *t* of its $N_t$ tweets have at least *t* retweets each and the other ($N_t$-*t*) tweets have ≤*t* retweets each. The tweets which are considered in the calculation of the t factor contain a link to a single publication, the publications of a researcher, the publications of a research group, etc. – depending on the level of analysis.

We deem the calculation of the t factor for publications appropriate because tweets cannot be treated as substantial publications producing their own impact. Although tweets are published material of authors, they are hints to other interesting objects (publications) rather than independent material being of significance. According to Bik and Goldstein (2013), "because Twitter serves as an information filter for many scientists, publicizing articles on social media can alert researchers to interesting studies that they may not have otherwise come across (e.g., research in journals tangential to their field or within discipline publications they do not normally read)". Thus, the allocation of impact to a certain publication measured by re-tweets via tweets (based on the t factor definition) makes sense.

**RESULTS**

Table 1 shows the authors of the tweets and re-tweets regarding the paper by Hirsch (2005) and its date of publication. Further, the tweets are categorized whether they are tweets or re-tweets. The column "number of re-tweets" shows the number of re-tweets for a single tweet in the subsequent line (the lines presenting the number of re-tweets are marked in grey). For example, the tweet published by "Biblioteca HUVH. ICS" (see line no. 5 in Table 1) was retweeted by "Biblioteca de Salut" two days later. It was followed by a second retweet by "Adela Zambrano" on June 24, 2012. Of the total n=69 tweets, n=28 are original tweets and n=41 are retweets. The most retweeted tweet has 25 retweets.

Table 1: Tweets and retweets regarding the paper by Hirsch (2005)

| No. | Author | Date | Tweet | Retweet | Number of retweets |
|---|---|---|---|---|---|
| 1 | Western Hammer | 17.06.2011 | 1 | | 0 |
| 2 | Osmar Arouck | 02.10.2011 | 1 | | 0 |
| 3 | Jonathan Jones | 09.03.2012 | 1 | | 0 |
| 4 | UniverEdu-KNEU | 07.06.2012 | 1 | | 0 |
| 5 | BibliotecaHUVH. ICS | 20.06.2012 | 1 | | |
| 6 | Biblioteca de Salut | 22.06.2012 | | 1 | |
| 7 | Adela Zambrano | 24.06.2012 | | 1 | 2 |
| 8 | JovenesSEF | 21.07.2012 | 1 | | 0 |
| 9 | CNB | 21.07.2012 | 1 | | |
| 10 | Laura F^2 | 21.07.2012 | | 1 | |
| 11 | Eduardo Oliver | 21.07.2012 | | 1 | |
| 12 | Sheila González | 21.07.2012 | | 1 | |
| 13 | SusRodriguezNavarro | 21.07.2012 | | 1 | |
| 14 | Andrea Cortés | 21.07.2012 | | 1 | |
| 15 | David Ochoa | 28.08.2012 | | 1 | 6 |
| 16 | Gil Kosgei | 28.08.2012 | 1 | | 0 |

| 17 | Nottingham research | 09.07.2013 | 1 |   | 0 |
|----|---------------------|------------|---|---|---|
| 18 | Green Ink | 09.07.2013 | 1 |   | 0 |
| 19 | Eric Michael Johnson | 09.07.2013 | 1 |   |   |
| 20 | Dr Mel Thomson | 09.07.2013 |   | 1 |   |
| 21 | Graham Steel | 09.07.2013 |   | 1 |   |
| 22 | Dr Simon Wells | 09.07.2013 |   | 1 |   |
| 23 | Dr. Christie Wilcox | 09.07.2013 |   | 1 |   |
| 24 | Anthony J. Martin | 09.07.2013 |   | 1 |   |
| 25 | Diana May | 09.07.2013 |   | 1 |   |
| 26 | Glenn Carlson PE PhD | 09.07.2013 |   | 1 |   |
| 27 | Britt Jeye | 09.07.2013 |   | 1 |   |
| 28 | Darren Milligan | 09.07.2013 |   | 1 |   |
| 29 | Nikolay Vavilov | 09.07.2013 |   | 1 |   |
| 30 | Lorna Quandt | 09.07.2013 |   | 1 |   |
| 31 | Sarah | 09.07.2013 |   | 1 |   |
| 32 | sooike stoops | 09.07.2013 |   | 1 |   |
| 33 | g.e@UBC | 09.07.2013 |   | 1 |   |
| 34 | Vince Tingey | 09.07.2013 |   | 1 |   |
| 35 | Sarah Pohl | 11.07.2013 |   | 1 |   |
| 36 | Jerrie Lynn Morrison | 13.07.2013 |   | 1 |   |
| 37 | Dr Raul Pacheco-Vega | 13.07.2013 |   | 1 |   |
| 38 | Rodrigo Nieto-Gomez | 13.07.2013 |   | 1 |   |
| 39 | David Moscrop | 13.07.2013 |   | 1 |   |
| 40 | yourqueerprof | 13.07.2013 |   | 1 |   |
| 41 | Michael Hiatt, PhD | 13.07.2013 |   | 1 |   |
| 42 | Sophien Kamoun | 13.07.2013 |   | 1 |   |
| 43 | Sarah Unruh | 13.07.2013 |   | 1 |   |
| 44 | RGWhite | 13.07.2013 |   | 1 | 25 |
| 45 | Heriberto Franco | 19.07.2013 | 1 |   | 0 |
| 46 | Heriberto Franco | 19.07.2013 | 1 |   | 0 |
| 47 | Heriberto Franco | 19.07.2013 | 1 |   | 0 |
| 48 | Heriberto Franco | 19.07.2013 | 1 |   | 0 |
| 49 | Heriberto Franco | 19.07.2013 | 1 |   | 0 |
| 50 | Heriberto Franco | 19.07.2013 | 1 |   | 0 |
| 51 | Benjamin Fox | 01.10.2013 | 1 |   |   |
| 52 | Pierre-Antoine Laloë | 01.10.2013 |   | 1 | 1 |
| 53 | Huntsman Library | 10.01.2014 | 1 |   |   |
| 54 | Snow College News | 10.01.2014 |   | 1 | 1 |
| 55 | Steve Royle | 20.05.2014 | 1 |   | 0 |
| 56 | David Schoppik | 10.06.2014 | 1 |   |   |
| 57 | Drug Monkey | 10.06.2014 |   | 1 | 1 |
| 58 | David Schoppik | 10.06.2014 | 1 |   | 0 |
| 59 | Drug Monkey | 10.06.2014 | 1 |   | 0 |
| 60 | Realscientists | 02.07.2014 | 1 |   | 0 |
| 61 | Nicholas A. Peppas | 09.09.2014 | 1 |   |   |
| 62 | Elif | 10.09.2014 |   | 1 |   |
| 63 | M N V Ravi Kumar | 10.09.2014 |   | 1 |   |
| 64 | pw_research_group | 10.09.2014 |   | 1 |   |

| 65 | Yvonne Perrie | 10.09.2014 |   | 1 | 4 |
|----|---------------|------------|---|---|---|
| 66 | Víctor Yepes | 19.02.2015 | 1 |   | 0 |
| 67 | Catherine Cottone | 02.04.2015 | 1 |   | 0 |
| 68 | Kinaba | 16.06.2015 | 1 |   |   |
| 69 | Yowa | 16.06.2015 |   | 1 | 1 |
|    | Total |            | 28 | 41 |   |

Table 2 shows the calculation of the t factor which is based on the column "number of retweets" in Table 1. As the table shows, there are 28 tweets with 0 to 25 retweets. Based on the t factor definition, we calculate a t factor of t=3 for the paper by Hirsch (2005): There are three tweets with t≥3 retweets. Those three tweets are part of the t core in analogy to the h core (Bornmann and Daniel, 2007). The t factor seems to be a very low number for this important paper, but one should consider that the paper has already been published in 2005 and Scopus only offers Twitter data for the time period between June 2011 and June 2015.

Table 2: Calculation of the t factor for the paper by Hirsch (2005)

| Tweetrank | Number of retweets | t core tweets |
|-----------|--------------------|---------------|
| 1 | 25 | 1 |
| 2 | 6 | 2 |
| 3 | 4 | 3 |
| 4 | 2 | - |
| 5 | 1 | - |
| 6 | 1 | - |
| 7 | 1 | - |
| 8 | 1 | - |
| 9 | 0 | - |
| 10 | 0 | - |
| 11 | 0 | - |
| 12 | 0 | - |
| 13 | 0 | - |
| 14 | 0 | - |
| 15 | 0 | - |
| 16 | 0 | - |
| 17 | 0 | - |
| 18 | 0 | - |
| 19 | 0 | - |
| 20 | 0 | - |
| 21 | 0 | - |
| 22 | 0 | - |
| 23 | 0 | - |
| 24 | 0 | - |
| 25 | 0 | - |
| 26 | 0 | - |
| 27 | 0 | - |
| 28 | 0 | - |

## DISCUSSION

Based on the definition of the well-known h index we propose the t factor for measuring the impact of publications on Twitter. The new index combines tweet and retweet data in a balanced way whereby retweets are seen as data reflecting the impact of initial tweets. The t factor is conceptualized similarly to the single publication h index. This index was introduced by Schubert (2009) for assessing single publications by using the times cited data for each citing publication to calculate the index. The single publication h index does not only consider the direct impact of publications, but also their indirect impact: "Citation indicators usually measure the 'direct impact' of publications, i.e., the amount of the citations received (whether in the form of simple counts, weighted sums or normalised units). Undoubtedly, however, publications may exert influence also indirectly, e.g., through their presence in reference lists. It seems reasonable to construct indicators that take into account not only the direct, but also the indirect citation influence of publications" (Schubert 2009, p.560). Also, the t factor does not only measure the direct impact of publications (by counting tweets), but also their indirect impact (by counting retweets separately). It measures how interesting single tweets were on publications for other Twitter users.

In this study, we have shown how the t factor for single publications is calculated. t factors for units in science with more than one publication (e.g. journals, single researchers, institutions) can be calculated in two different ways: (i) the tweets and retweets of all publications published by the unit are used for the t factor calculation – independently of the tweets and retweets belonging to single publications; (ii) another way would be to calculate the t factor for each single publication and to calculate a mean value or a second-order t factor over the single t factor values. The second order t factor is a factor which is based on the first-order t factor values. It should be the tasks of future studies to ascertain which way of calculation should be preferred (see here also the proposals of Egghe 2011a, 2011b).

The t factor has several advantages (against bare Twitter counts): (i) tweets and retweets are used as different impact information for a publication. The differentiation is necessary, because retweets are only repetitions of initial tweets and show their success at the followers; (ii) The example in this study shows that Twitter data is skewed – similar to citation data. The h index definition (adapted to Twitter data) is able to handle skewed data in a balanced way so that tweets with an enormous number of retweets do not distort the result; (iii) the factor is very simple to calculate and could be included in databases like Scopus. Also, the t factors for single publications can be used to calculate percentiles for a subject category or time period (Bornmann und Haunschild 2016). Then, the percentile impact value for a single publication could be presented in Scopus; (iv) the t factor has been developed against the backdrop of a proved index proposal in bibliometrics. Bornmann et al. (2011) investigated the quality of the single publication h index, and Egghe (2010) its functional relation with other metrics. Further, a web application has been programmed to calculate this index for selected publications (Thor and Bornmann 2011).

## CONCLUSIONS

Twitter data is an interesting data source for measuring the impact of publications on Twitter: "Citations from Twitter are a particularly interesting data source, since they capture the sort of informal discussion that accompanies early important work. There is, encouragingly, evidence that Tweeting scholars take citations from Twitter seriously, both in creating and reading them" (Priem 2014). In this study we have introduced the t factor to measure impact on Twitter in a balanced way: A unit (single publication, journal, researcher, research group etc.) has factor *t* if *t* of its $N_t$ tweets have at least *t* retweets each and the other ($N_t$-t) tweets have ≤*t* retweets each. The t factor combines the number of tweets (for a publication, researcher etc.) and the number of retweets for the single tweets. We encourage future studies to investigate the mathematical properties and the empirical quality of the t factor.

Also, studies could check whether it is sensible to combine the number of tweets with the number of times each tweet was favored (instead of its number of retweets). Such an *f* factor would be defined as follows: A unit (single publication, journal, researcher, research group etc.) has factor *f* if *f* of its *Nf* tweets were at least *f* times favored each, and the other (*Nf-f*) tweets were ≤ *f* times favored each. Of course, variants of t factor and f factor could be calculated also for other altmetric sources (e.g. Facebook) where posts are repeated or favored (liked, shared, etc.) by other users.

## ACKNOWLEDGEMENT

This research received no specific grant from any funding agency in the public, commercial, or not-for-profit sectors.